\documentclass[11pt,a4paper]{article}
\usepackage[T1]{fontenc}
\usepackage[utf8]{inputenc}
\usepackage{authblk}
\usepackage{graphicx}

\usepackage{subfig}
\usepackage{amsmath}
\newcommand{\un}[1]{\ensuremath{\mathrm{\,#1}}}

\title{Optimal estimation of entanglement and discord in two-qubit states}

\author[1,2]{Salvatore Virzì}
\author[1,3]{Enrico Rebufello}
\author[1,*]{Alessio Avella}
\author[1]{Fabrizio Piacentini}
\author[1]{Marco Gramegna}
\author[1]{Ivano Ruo Berchera}
\author[1]{Ivo Pietro Degiovanni}
\author[1]{Marco Genovese}
\affil[1]{INRIM, Strada delle Cacce 91, 10135 Torino, Italy}
\affil[2]{Universit\`a degli Studi di Torino, Dipartimento di Fisica, Via Giuria 1, 10125 Torino, Italy}
\affil[3]{Politecnico di Torino, Corso Duca degli Abruzzi 24, I-10129 Torino, Italy}

\affil[*]{a.avella@inrim.it}

\begin{document}

\flushbottom
\maketitle

\begin{abstract}
	
	Recently, the fast development of quantum technologies led to the need for tools allowing the characterization of quantum resources. In particular, the ability to estimate non-classical aspects, e.g. entanglement and quantum discord, in two-qubit systems, is relevant to optimise the performance of quantum information processes. Here we present an experiment in which the amount of entanglement and discord are measured exploiting different estimators. Among them, some will prove to be optimal, i.e., able to reach the ultimate precision bound allowed by quantum mechanics. These estimation techniques have been tested with a specific family of states ranging from nearly pure Bell states to completely mixed states. This work represents a significant step in the development of reliable metrological tools for quantum technologies.

\end{abstract}

\section*{Introduction}

The problem to quantifying the amount of quantum resources in physical systems is strongly acknowledged by the physicists community, both for applications concerning quantum information technologies and experiments on quantum mechanics foundations 

The reconstruction of the density matrix, by means of the quantum state tomography, provides all the information on the physical system under analysis \cite{tomo1,tomo2}. However, quantum state tomography is a demanding procedure in terms of quantum resources due the high number of measurements required on identical copies of the system. Moreover, it has two main limitations that could be critical for several applications: First of all, reconstructions are based on optimisation algorithms applied to likelihood functions, therefore, a tomography does not allow to perform an easy estimation of the uncertainty associated to the reconstructed density matrix. On the other hand, quantum state tomography becomes impractical for high-dimensional systems \cite{unc1,unc2}. In addition, a full knowledge of the density matrix does not provide an immediate quantification of the amount of the quantum resource needed, hence, it is necessary introduce dedicated parameters.

Among the most relevant and exploited quantum resources, a crucial role is played by entanglement and discord, whose estimation is of the utmost relevance for present and upcoming quantum technologies. In general, the parameters used to evaluate them are defined for well specific families of quantum states, and several measurements have to be performed in order to experimentally obtain their values. 

In particular, the measurement of the amount of entanglement is a parameter estimation problem where the value of entanglement is obtained indirectly from the measurement of one or more proper observables. A quantitative measure of entanglement corresponds to a non-linear function of the density operator, and it is not possible to identify a quantum observable directly associated to it.
Several theoretical and experimental works have addressed this topic \cite{rev1,rev2,rev3,rev4}, providing different approaches to efficiently estimate the amount of entanglement of a quantum state from a reduced set of measurements \cite{appro1,appro2,appro3,appro4,appro5}, e.g. visibility measurements \cite{vis}, Bell tests \cite{bell1}, entanglement witnesses \cite{w1,w2,w3,w4,w5}, or are related to Schmidt number \cite{based10,based11,based12}. Many of these techniques have also been implemented in laboratory \cite{lab1,lab2,lab3,lab4,lab5,lab6,lab7,deco1,deco2}. 

Quantum discord, instead, is a figure of merit that can be used to quantify non-classicality of correlations within a physical system \cite{D1,D3,D5,D4,D6,PhysRevA.87.052136,Dakic2012}. 
Separability of the density matrix describing a multi-partite state does not guarantee vanishing of the discord, demonstrating that absence of entanglement does not imply classicality. Quantum discord has been proposed as the key resource needed for certain quantum communication tasks and quantum computational models not entirely relying on entanglement. Due to the high interest on quantum discord, both for foundational aspects of quantum mechanics and for applications, techniques allowing to estimate this quantity are demanded. 
%%%%%%%%%%%%%%%%%%%%%%%%%%%
% Nuovo pezzo
%%%%%%%%%%%%%%%%%%%%%%%%%%%
Unfortunately, in general, the discord doesn't present an analytical expression. Therefore, we take in account a geometrical approximation\cite{PhysRevA.83.052108} for our extimation task.
%Particularly relevant for this issue is the geometric measure of quantum discord \cite{QGD1,QGD2}.  

In many applications, specially for the quantum information technologies,a robust and resource-efficient protocol to estimate such quantities is highly demanded. Therefore, the optimisation problems concerning the ultimate precision bounds on entanglement and discord, and the optimal measurements achieving those bounds have been investigated\cite{brida,brida2}. That procedure is self-consistent and allows reaching the ultimate precision imposed by the quantum Cram{\'e}r-Rao bound \cite{fisher1}, i.e. the minimum theoretical uncertainty compatible with the local quantum estimation theory \cite{theory1,theory2,theory3,theory4,theory5}, obtained by maximizing the Fisher Information \cite{fisher1,fisher2}. 

Here, we exploit three different parameters\cite{negconc} providing quantitative information on the amount of  entanglement in qubit states: Negativity, Log-Negativity and Concurrency. For each of these parameters, we introduce two different estimators: one non-optimal and one optimal (i.e. saturating the quantum Cramér-Rao bound).
In addition to entanglement, we also introduce an optimal procedure to estimate Quantum Geometric Discord\cite{QGD1, QGD2}, providing the best analytical approximation of the amount of quantum discord for the family of states under exam (defined below). 
This effort represents a sharp advancement with respect to our previous work\cite{brida,brida2}, since here we extend the entanglement estimation analysis to different parameters (Log-Negativity and Concurrence) and we addressing for the first time optimal estimation quantum discord.

The paper is organised as follows: first of all, we introduce the estimators and related precision bounds obtained in agreement to quantum estimation theory. Then, we describe our experiment aiming to estimate the amount of entanglement and discord of a large class of two photon states. Finally, we compare experimental results, and their related uncertainty, with the theoretically-expected ones.

\section*{Estimators definition}

%$Variance \geq \mathcal{J}$, where $\mathcal{J}$ is the Quantum Fisher Information [...].

We consider four different parameters:  Negativity, Log-Negativity, Concurrency and Quantum Geometric Discord, allowing, to quantify the amount of entanglement or discord in two-qubit systems.
For each parameter, we introduce two estimators, one optimal and one non-optimal, allowing to estimate it with a lower number of measurements with respect to a full reconstruction of the density matrix. However, to define such estimators, we need some \textit{a priori} knowledge of the family of quantum systems we are going to test. In particular, our estimators are suited for quantum states whose density matrix can be expressed in the following form:
\begin{equation}
\rho =(1-p) \left(
\begin{array}{cccc}
0 & 0 & 0 & 0 \\
0 & 1/2 & 0 & 0 \\
0 & 0 & -1/2 & 0 \\
0 & 0 & 0 & 0 \\
\end{array}
\right)+p \left(
\begin{array}{cccc}
0 & 0 & 0 & 0 \\
0 & q & -\sqrt{q (1-q)} & 0 \\
0 & -\sqrt{q (1-q)} & 1-q & 0 \\
0 & 0 & 0 & 0 \\
\end{array}
\right)
\label{state}
\end{equation}
where $p$ and $q$ are unknown variables within the interval [0,1]. This includes states with different entanglement amount, ranging from the singlet state (maximally entangled) to a completely de-coherent mixture. These are typical quantum states involved in many real scenarios in which the entangled qubits are exposed to decoherence due to coupling with the environment, degradating the quantum resources available for the task we want to use them for, thus making them particularly worth investigating.

In the following, for each parameter, we define the estimators and we calculate the corresponding theoretical minimal uncertainty.

\subsection*{Negativity}  Negativity of entanglement is defined by:
\begin{equation}
\mathcal{N}=|| \rho^{T_A}|| -1  ,
\label{e1}
\end{equation}
where: $\rho^{T_A}$ is the partial transpose of $\rho$ with respect to the subsystem $A$ and $||X||=
Tr \sqrt{X^\dagger X} $ is the trace norm of the operator $X$. Negativity ranges from 0 to 1, where 1 is the negativity of a maximally entangled states and 0 is the negativity of a completely separable states.
For the family of states taken into account (Eq. \ref{state}), the negativity becomes:
\begin{equation}
\mathcal{N}=  2 p \sqrt{q (1-q)}.
\label{e2}
\end{equation}

Exploiting the Quantum Fisher Information it is possible calculate the quantum Cram{\'e}r-Rao bound for the estimation of the negativity: 
\begin{equation}
QCRB_{\epsilon \mathcal{N}} =1-\mathcal{N} ^2,
\label{v3}
\end{equation}
representing the minimum variance obtained for the estimation of negativity in a single measurement. Thus, the optimal estimation of the negativity has the uncertainty:
\begin{equation}
unc_{Opt\epsilon\mathcal{N}} = \sqrt{QCRB_{\epsilon \mathcal{N}}} =\pm\sqrt{1-\mathcal{N} ^2}
\label{e3}
\end{equation}

We define a non-optimal estimator $\epsilon \mathcal{N}_1$:
\begin{equation}
\epsilon \mathcal{N}_1 = 1 -4  P( ++ ) ,
\label{neg1}
\end{equation}
where $P(x)$ is the probability of the event $X$ and, here as well as in the following, the symbol +(-) indicates projection onto the state $|+(-)\rangle=\frac{|H\rangle +(-) |V\rangle}{\sqrt2}$.
The theoretical minimum uncertainty associated to the non-optimal estimator $\epsilon \mathcal{N}_1$ is:
\begin{equation}
unc_{\epsilon \mathcal{N}_1} = \pm \sqrt{-\left(\mathcal{N} ^2+2 \mathcal{N} -3\right)}.
\label{e4}
\end{equation}

Then, we define an optimal estimator $\epsilon \mathcal{N}_2$:
\begin{equation}
\epsilon \mathcal{N}_2 =  P( +- ) + P( -+ ) - P( ++ ) - P( -- ),
\label{neg2}
\end{equation}
with the associated theoretical minimum uncertainty corresponding to the one set by the saturation of the quantum Cram{\'e}r-Rao bound (Eq. \ref{e3}).

\subsection*{Log-Negativity}  

This parameter is defined as:
\begin{equation}
\mathcal{L}= \log_2{( || \rho^{T_A}|| )  }.
\label{l1}
\end{equation}
For the family of states taken into account, the Log-Negativity can be expressed as:
\begin{equation}
\mathcal{L}= \log _2(2 p \sqrt{q (1-q)}+1).
\label{l2}
\end{equation}
The corresponding quantum Cram{\'e}r-Rao bound is:
\begin{equation}
QCRB_{\epsilon \mathcal{L}} =  -\frac{2^{-\mathcal{L}} \left(2^{\mathcal{L}}-2\right)}{\log ^2(2)}
\label{QCRBlog1}
\end{equation}
We define the non-optimal estimator $\epsilon \mathcal{L}_1$:
	\begin{equation}
\epsilon \mathcal{L}_1 = \log _2\left(1-4 \left(P\left(+,+\right)-\frac{1}{4}\right)\right)
\label{l3}
\end{equation}
having the following minimum uncertainty:
\begin{equation}
unc_{\epsilon \mathcal{L}_1} = \pm \sqrt{-\frac{4^{-\mathcal{L}} \left(4^{\mathcal{L}}-4\right)}{\log ^2(2)}}
\label{l6}
\end{equation}
Moreover, we define the optimal estimator $\epsilon \mathcal{L}_2$:
	\begin{equation}
\epsilon \mathcal{L}_2 = \log _2\left(1-\left(P\left(-,-\right)-P\left(-,+\right)-P\left(+,-\right)+P\left(+,+\right)\right)\right).
\label{l4}
\end{equation}
The theoretical uncertainty associated to this estimator corresponds to the square root of the quantum Cram\'er-Rao bound (Eq. \ref{QCRBlog1}) for the Log-Negativity:
\begin{equation}
	unc_{\epsilon\mathcal{L}_2} = \pm \sqrt{-\frac{2^{-\mathcal{L}} \left(2^{\mathcal{L}}-2\right)}{\log ^2(2)}}
	\label{log1}
\end{equation}

\subsection*{Concurrence}

Concurrence is defined as:
	\begin{equation}
	\mathcal{C}=max(0, \lambda_1 - \lambda_2 - \lambda_3 - \lambda_4),
	\end{equation}
	where $\lambda_i$ are eigenvalues of the matrix $R = \sqrt{ \sqrt{\rho} (\sigma_y \otimes \sigma_y) \rho^\ast (\sigma_y \otimes \sigma_y) \sqrt{\rho}}$ in descending order, and $\sigma_y$ is the Pauli matrix $\begin{bmatrix}
0       & i   \\
	-i     &0  \\
	\end{bmatrix}$.
	For the family of states described by Eq. \ref{state}, the concurrence become:
	\begin{equation}
	\mathcal{C}=  2 p \sqrt{q (1-q)}.
	\end{equation}
Having Concurrence and Negativity the same theoretical value, for the family of states taken into account, we can use the same estimators previously introduced in Eq.s \ref{neg1} and \ref{neg2}.
	
	\subsection*{Quantum Geometric Discord}
		
As previously stated, we are also interested in the amount of discord of a state. In order to present a valid estimation technique for all the bipartite states represented by Eq. \ref{state}, we use the Quantum Geometric Discord ($\mathcal{Q}$). This geometrical approximation is the best indicator for the amount of the discord in the states under exam. The corresponding quantum Cram{\'e}r-Rao bound is:
\begin{equation}
QCRB_{\epsilon \mathcal{Q}} = 2 (1-2 \mathcal{Q} ) \mathcal{Q}.
\label{QCRBq3}
\end{equation}

Both the non-optimal estimator $\epsilon \mathcal{Q}_1$ and the optimal one $\epsilon \mathcal{Q}_2$ are functions, respectively, of the estimators $\epsilon \mathcal{N}_1$ and $\epsilon \mathcal{N}_2$ defined in Eq.s \ref{neg1} and \ref{neg2}:	
	\begin{equation}
	\epsilon \mathcal{Q}_1 = \frac{1}{2}\left(\epsilon \mathcal{N}_1 \right)^2,\\ 
	\epsilon \mathcal{Q}_2 = \frac{1}{2}\left(\epsilon \mathcal{N}_2 \right)^2.
	\label{q1}
	\end{equation}
	The theoretical uncertainty associated to the non-optimal estimator $\epsilon \mathcal{Q}_1$ is:
	\begin{equation}
	unc_{\epsilon \mathcal{Q}_1} = \pm \sqrt{-2 \mathcal{Q}  \left(2 \mathcal{Q} +2 \sqrt{2} \sqrt{\mathcal{Q} }-3\right)}.
	\label{q4}
	\end{equation}
	The theoretical uncertainty associated to the optimal estimator $\epsilon \mathcal{Q}_2$ is the one saturating the quantum Cram{\'e}r-Rao bound (Eq. \ref{QCRBq3}).
	\begin{equation}
	 unc_{\epsilon \mathcal{Q}_2} = \pm \sqrt{QCRB_{\epsilon \mathcal{Q}}}.
	\label{q5}
	\end{equation}

\section*{Experimental apparatus}

The family of entangled states investigated in our work, is constituted by two photons polarization-entangled states obtained exploiting the phenomenon of spontaneous parametric down conversion (SPDC). 

\begin{figure}[ht!]
	\centering
	\includegraphics[width=12cm]{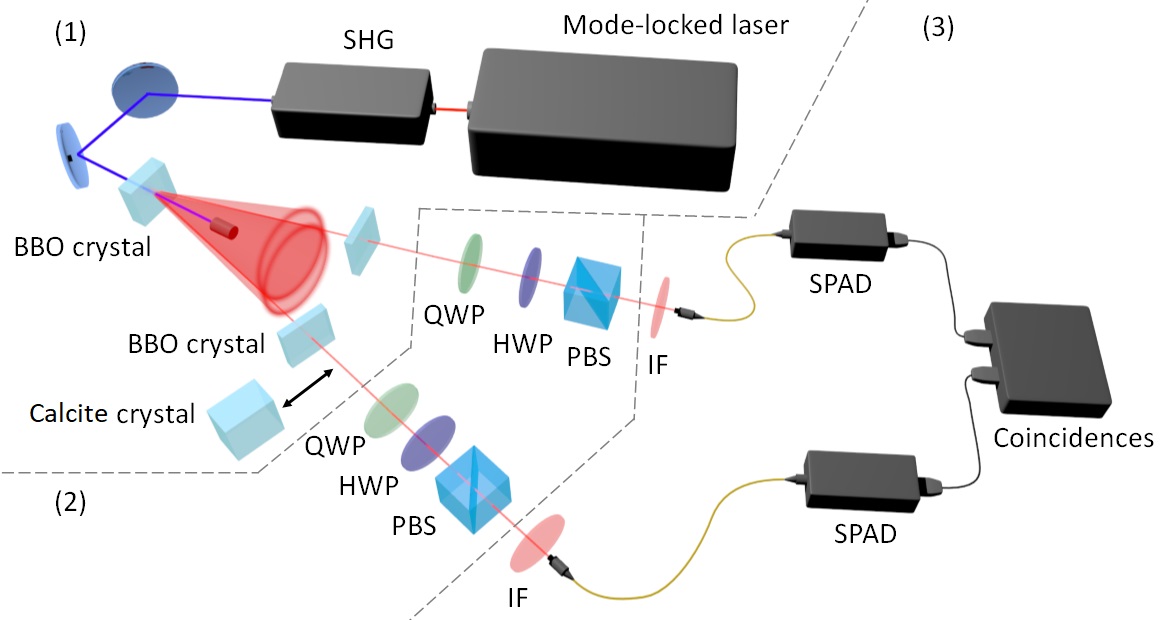}
	\caption{The experimental apparatus can be divided in three parts.  (1) A source of polarisation entangled photons, based on SPDCs, composed by: a Ti:sapphire mode-locked laser, a second harmonic generator (SHG), a BBO non-linear crystal for SPDC generation and two BBO crystals for walk-off compensation. A $2.7\un{mm}$ length birefringent crystal, with optical axis orthogonal to the photon propagation direction, can be inserted in one of the paths in order to introduce decoherence. (2) A tomographic measurement apparatus, composed for each path of: a quarter wave plate (QWP), a half wave plate (HWP) and a polarising beam splitter (PBS). (3) The detection system, comprising two interference filters (IF), two fiber couplers injecting the photons into multi-mode fibers addressing them to Silicon single-photon avalanche diodes (SPADs), and coincidence electronics.}
	\label{fig:setup}
\end{figure}

The first part of the set-up (corresponding to the region (1) in Fig. \ref{fig:setup}) is a source of polarization-entangled photons based on a scheme \cite{kwiat} exploited in many experiments concerning foundation of quantum mechanics and quantum technologies \cite{lab1}. In particular, our scheme is based on a Ti:Sapphire mode-locked laser, emitting pulses with duration of $150\un{fs}$ at a wavelength centred on $808\un{nm}$. Such laser beam induces the second harmonic generation in a lithium triborate (LBO) non-linear crystal. The resulting beam, with a central wavelength at $404\un{nm}$, is used to pump a $0.5\un{mm}$ long $\beta$-barium borate (BBO) non-linear crystal where type II SPDC process occurs, generating correlated photon pairs \cite{nist}. Two irises are used to spatially select the photons belonging to the intersections of the horizontally- and vertically-polarized degenerate SPDC cones ($808\un{nm}$). On each of the two selected paths, a $0.25\un{mm}$ thick BBO crystal is used to compensate the temporal delay between the horizontally and the vertically polarised photons induced by the birefringence within the SPDC crystal. At the output of these crystals, ideally, the polarisation-entangled photons are in the state:
\begin{equation}
| \psi_{  \phi } \rangle = \frac{ | HV \rangle +  e^{i \phi}| VH \rangle}{\sqrt{2}}
\end{equation}
(being H and V, respectively, the horizontal and vertical polarisation components), with a relative phase $\phi$ between the ordinary and extraordinary polarized light. A fine tilting of one of the compensation crystals is performed to tune the parameter $\phi$. 

It is possible to introduce decoherence in our entangled state by introducing, in one of the two paths, an additional birefringent crystal with an sufficient thickness (for this purpose we use a $2.7\un{mm}$ thick calcite crystal). 

The second part of the set-up (corresponding to the region (2) in Fig. \ref{fig:setup}) is a typical polarisation quantum tomographic apparatus \cite{geno}. Each path is equipped with a quarter wave plate (QWP), a half wave plate (HWP) and a polarising beam splitter (PBS), allowing to project each photon polarisation on any state of the Bloch sphere surface.

Finally (in the region (3) of Fig. \ref{fig:setup}), for each path, an interference filter (IF) spectrally selects the photons, subsequently injected into a multi-mode fibre. Then, the fibre sends the photons to a Silicon single-photon avalanche diode (SPAD) for the detection. A dedicated time correlated counting system is used to perform temporal post-selection on photon counts.

\section*{Results}

In our experiment, we produce the set of quantum states described in Eq. \ref{state} performing, in post processing, a statistical mixture of a physical pure singlet state $|\Psi_-\rangle$ and a completely decoherent mixture. For each state of our final set, we perform all the measurements, required to evaluate the estimators, and we calculate the associated statistical uncertainty. 

\begin{figure}[ht]
	\centering
	\includegraphics[width=12cm]{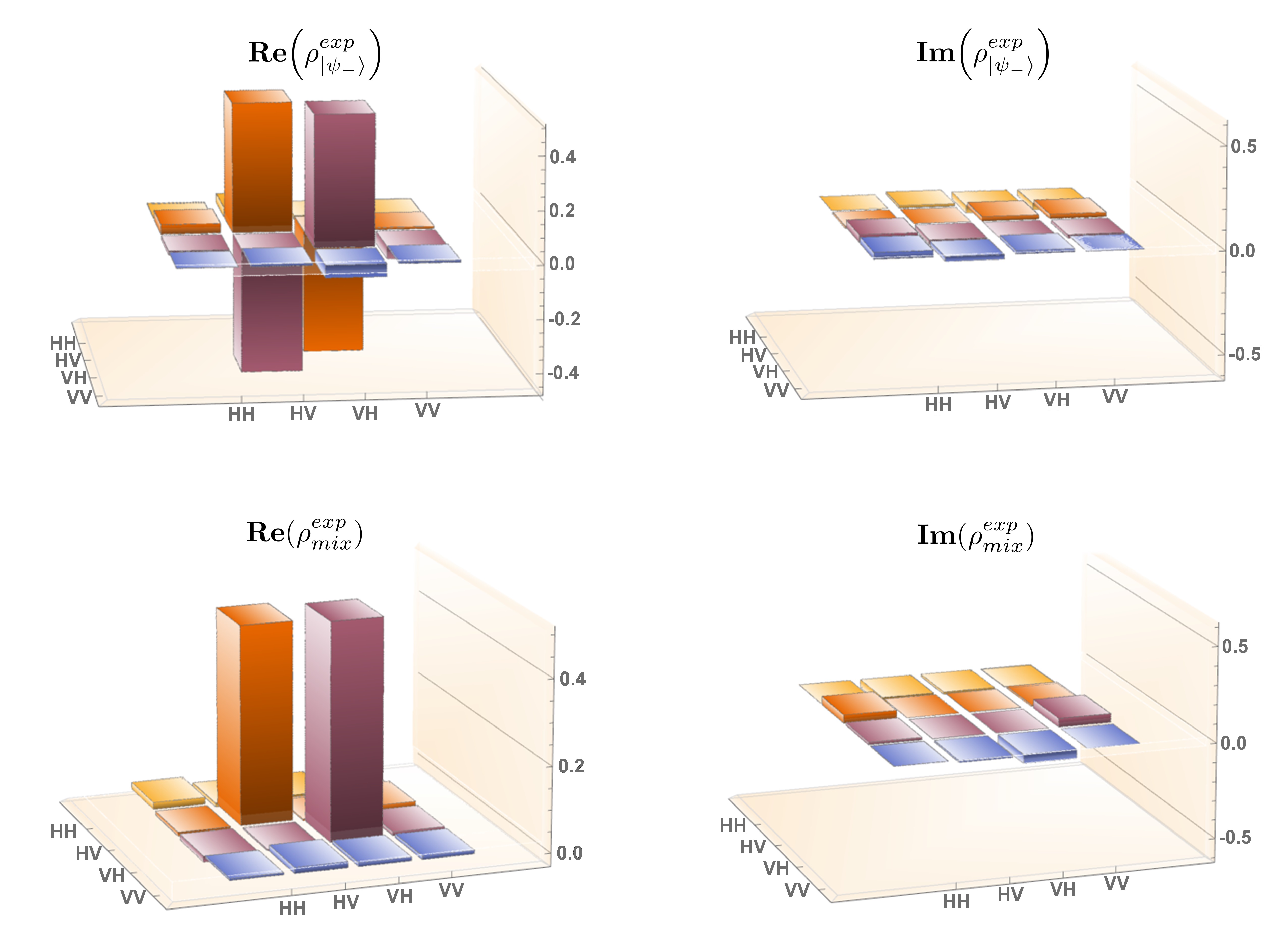}
	\caption{Real (left) and imaginary (right) part of the tomographically reconstructed density matrix for the singlet, maximally entangled, state (top) and the completely decoherent mixture (bottom).}
	\label{fig:tomo}
\end{figure}

To determine the quality of the states produced in our experiment, we exploit the quantum state tomography technique and we calculate the Uhlmann’s Fidelity \cite{Uhlmann} of the reconstructed state with respect to the theoretical expectations:
\begin{equation}
\mathcal{F} = Tr \left( \sqrt{\sqrt{\rho^{exp}} \rho^{th} \sqrt{\rho^{exp}}  }  \right).
\end{equation}
where, $\rho^{exp}$ is the reconstructed density matrix and $\rho^{th}$ is the corresponding theoretical one.
The experimentally reconstructed matrices of the singlet state and of the decoherent mixture generated in our setup are shown in Fig. \ref{fig:tomo}, while the corresponding theoretical matrices can be written respectively, in the H-V basis, as:
\begin{equation}
\begin{aligned}
\rho^{th}_{| \psi_{-} \rangle} = \begin{bmatrix}
0 & 0 & 0 & 0  \\
0 & 1/2 & -1/2 & 0  \\
0 & -1/2 & 1/2 & 0  \\
0 & 0 & 0 & 0 
\end{bmatrix} & & & & & & and & & & & & &
\rho^{th}_{| \psi_{mix} \rangle} = \begin{bmatrix}
0 & 0 & 0 & 0  \\
0 & 1/2 & 0 & 0  \\
0 & 0 & 1/2 & 0  \\
0 & 0 & 0 & 0 
\end{bmatrix},
\end{aligned}
\label{mat}
\end{equation}
where the choice to operate with a singlet states implies $q=1/2$ (see Eq. \ref{state}).

The values of Fidelity in our experiment are $\mathcal{F}_{\psi_-} = 0.975$ and $\mathcal{F}_{\psi_{mix}} = 0.985$ respectively. 

\begin{figure}[h!]
	\centering
	\includegraphics[width=12cm]{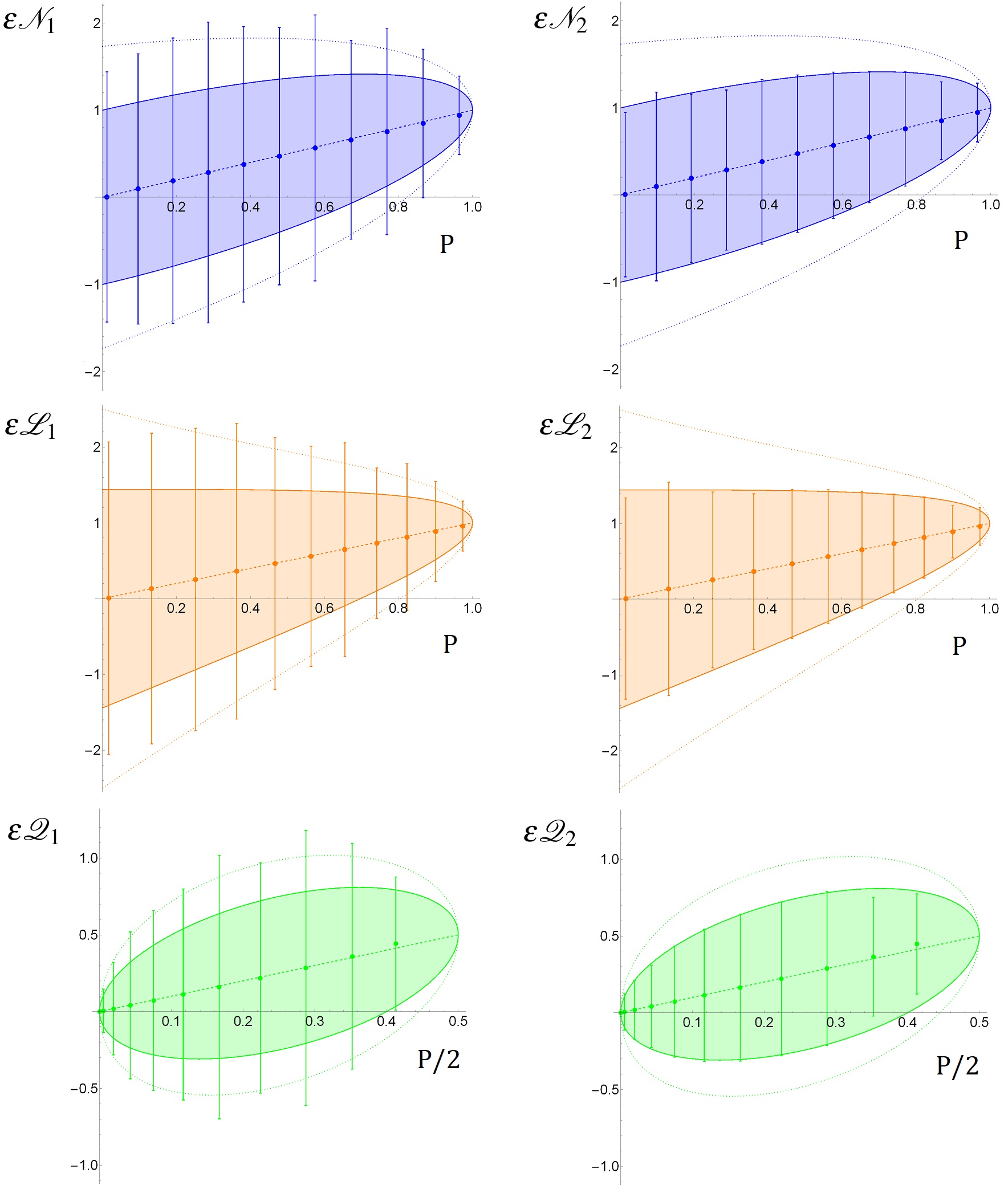}
	\caption{Results for Negativity (blue), Log-negativity and Concurrence (red) and Quantum Geometric Discord (green) non-optimal (left side) and optimal (right side) estimators, with respect to $p$ (see Eq.\ref{state}). Experimental points are compared with the following curves: theoretical value of the quantity to estimate (dashed curve), theoretical uncertainty for the non-optimal estimator (dotted curve) and the quantum Cram{\'e}r-Rao bound (solid curve).}
	\label{fig:neg}
\end{figure}

In Fig. \ref{fig:neg} are shown the main experimental results of this work. The experimental points concerning the several estimators introduced in this paper are plotted in function of the mixing parameter $p$ (defined in Eq. \ref{state}) ranging from $0$ (completely decoherent mixture) to $1$ (pure entangled state). For each point, the value of $p$ is evaluated exploiting the tomographical reconstruction of the density matrix of the corresponding quantum state. Each point results from the average on 10 independent estimations. The uncertainty bars associated with the experimental points represent the standard deviation of the measurement results statistical distribution, i.e. the statistical uncertainty associated with a single measurement. Experimental points are compared with the theoretical value of the estimator, represented by a dashed line. The experimental uncertainty bars are compared with the theoretical value of the uncertainty derived by the quantum Fisher information. Dotted lines represent the theoretical uncertainty for the non-optimal estimator, while solid lines indicate the theoretical uncertainty for the optimal estimator, i.e. saturating the quantum Cram{\'e}r-Rao bound, representing the minimum uncertainty allowed by quantum estimation theory.  All the theoretical curves, shown in Fig. \ref{fig:neg}, are calculated exploiting the knowledge of the experimental values of the parameters $p$ and $q$ obtained from the tomographical reconstruction of the density matrices (see Fig. \ref{fig:tomo}) of the physical systems involved in the experiment.

In Fig. \ref{fig:neg} different colours have been used for different parameters, in particular: blue for Negativity, orange for Log-negativity and green for Quantum Geometric Discord. On the left side of Fig. \ref{fig:neg} are shown the graphs concerning the non-optimal estimators for each parameters, while, on the right side are shown the optimal estimators plots. The plots show a good agreement between experimental results and theoretical predictions for each of the estimators, both for the value itself and the statistical uncertainty associated with it.
This is particularly relevant and interesting for the optimal estimators case, where our results demonstrate saturation of the Quantum Cram{\'e}r-Rao bound.

\newpage

\section*{Conclusion}

We performed an experiment comparing several non-classicality parameters related either to entanglement or discord. We directly extract the amount of entanglement with Negativity, Concurrence and Log-negativity, while we approximately evaluate the amount of discord by estimating the Quantum Geometric Discord. For each of these quantities we introduce two estimators, a non-optimal one and an optimal one, for a particular family of  states that have a recognised importance in the field of quantum information and related technologies. By evaluating the statistical uncertainties as the standard deviations of repeated measurements performed, we achieve a good agreement between the theoretical predictions and the experimental results. In particular, we demonstrate that predicted optimal estimators reach the ultimate theoretical precision limit represented by the quantum Cram{\'e}r-Rao bound. These results pave the way to diffusely use these estimators in quantifying resources for quantum technologies.

\section*{Acknowledgements}

This work has received funding from the European Union Horizon 2020 and the EMPIR Participating States in the context of the projects EMPIR-14IND05 'MIQC2', EMPIR-17FUN01  “Become”  and  EMPIR-17FUN06  “SIQUST”.

%\bibliography{sample}
%\bibliographystyle{plain}

\end{document}